\documentstyle [twocolumn,epsf]{mn}
\newcommand{\mincir}{\raise
-2.truept\hbox{\rlap{\hbox{$\sim$}}\raise5.truept\hbox{$<$}\ }}
\newcommand{\magcir}{\raise
-2.truept\hbox{\rlap{\hbox{$\sim$}}\raise5.truept\hbox{$>$}\ }}
\newcommand{\minmag}{\raise
-2.truept\hbox{\rlap{\hbox{$<$}}\raise6.truept\hbox{$<$}\ }}
\newcommand{\be}{\begin{equation}}
\newcommand{\ee}{\end{equation}}
\newcommand{\ba}{\begin{eqnarray}}
\newcommand{\ea}{\end{eqnarray}}
\newcommand{\brr}{\begin{array}}
\newcommand{\err}{\end{array}}
\newcommand{\bc}{\begin{center}}
\newcommand{\ec}{\end{center}}

\newcommand{\rosat}{{\it ROSAT~}}
\newcommand{\heao}{{\it HEAO-1~}}

\title[The \rosat X-ray Background Dipole]
     {The \rosat X-ray Background Dipole}
\author[M. Plionis \& I. Georgantopoulos]
       {M. Plionis \& I. Georgantopoulos \\
National Observatory of Athens, I. Metaxa \& B. Pavlou, Lofos Koufou, 
Palaia Penteli, 15236, Athens, Greece 
}

\begin{document}

\maketitle

\begin {abstract}
We estimate the dipole of the diffuse 1.5 keV X-ray background from 
the \rosat all-sky survey map of Snowden et al (1995).
We first subtract the diffuse  Galactic emission  by 
fitting to the data an exponential scale height, finite radius, disk model. 
 We further exclude regions of low galactic latitudes,
of local X-ray emission (eg the North Polar Spur) and model
 them  using two different methods.
We find that the \rosat X-ray background dipole points towards 
$(l,b) \approx (288^{\circ}, 25^{\circ}) \pm 19^{\circ}$  in consistency 
 with the Cosmic Microwave Background (within $\sim 30^{\circ}$); 
its direction is also in good agreement 
 with the \heao X-ray dipole at harder energies.  
 The normalised amplitude of the \rosat XRB dipole is  $\sim 1.7$ per cent.   
Subtracting from the \rosat map the expected X-ray background dipole 
due to the reflex motion of the observer with respect to the cosmic rest frame 
(Compton-Geting effect) we find the large-scale dipole of the X-ray emitting
extragalactic sources having an amplitude 
${\cal D}_{\rm LSS} \sim 0.9 {\cal D}_{\rm XRB}$, in general agreement 
with the predictions of Lahav et al (1997).
We finally estimate that the Virgo cluster is responsible for $\sim 20$ 
per cent of the total measured XRB dipole amplitude.

\vspace{5mm}


\end{abstract}

\section{Introduction}
According to the paradigm that the X-ray background (XRB) originates 
mainly from sources at redshifts $1 < z < 3$ (eg Shanks et al. 1991), 
 it should provide a means of measuring the well 
established solar motion with respect to the cosmic rest-frame, defined by the
Cosmic Microwave Background (CMB), towards $l=264^{\circ}$, $b=48^{\circ}$ (cf.
Lineweaver et al 1996). An imprint of this reflex motion should be 
a dipole pointing towards the direction of the CMB dipole (Compton-Getting 
effect). The available all-sky X-ray maps (\heao and \rosat) 
are composed by X-ray counts not only originating from such 
distant sources but also from local extragalactic sources ($z<0.1$). 
 These gravitating extragalactic objects, which cause our peculiar
motion with respect to the cosmic rest frame, emit X-rays and therefore their 
dipole should also point towards the CMB dipole. 
This has been demonstrated for the case of AGNs (Miyaji \& Boldt 1994) and 
of X-ray clusters (Lahav et al 1989; Plionis \& Kolokotronis 1998).
Therefore, the dipole pattern of the XRB results from at least two effects;
(a) the motion of the observer with respect to the XRB which in the
context of the cosmic-ray background was discussed first by Compton \& Getting 
(1935) and (b) the X-ray emission of extragalactic objects the gravitational 
field of which causes the observers motion with respect to the cosmic rest 
frame. The relative contribution of these effects have been analytically 
estimated by Lahav, Piran \& Treyer (1997) and were found to be of the same 
order of magnitude.

The dipole anisotropy of the XRB has been measured in hard X-rays 
(2-10 keV) using 
{\small UHURU} (Protheroe, Wolfendale \& Wdoczyk 1980) and  \heao 
 data  (Shafer \& Fabian 1983,  Jahoda 1993). The  
 resulting dipole was found to point towards the general direction of 
the CMB dipole but with a large uncertainty:  the 90 per cent 
confidence levels quoted by Shafer \& Fabian (1983) cover  12 per cent 
of the whole sky. 
 The use of the \rosat all-sky survey can extend 
these studies to soft energies with higher sensitivity 
and angular resolution.  
Kneissl et al. (1997)  presented a cross-correlation 
of the {\small COBE} DMR and the \rosat all-sky survey maps. 
However, their attempt to measure the extragalactic dipole was hampered by the 
contamination of the Galactic emission 
which is appreciable, especially 
at low Galactic latitudes, even in the hard \rosat band. They concluded that  
proper modelling of the Galactic contribution is necessary 
 in order to obtain a measurement of the dipole.   
In this letter,  we  attempt to model the diffuse 
Galactic component using a simple disk model.  
 After subtracting our Galaxy model, we estimate the cosmological
X-ray dipole using the \rosat all-sky survey hard maps (1.5 keV)
of Snowden et al. (1995).

\section{The \rosat all-sky survey data}
We use the \rosat all-sky survey maps at a mean energy 
of 1.5 keV (PSPC PI channels 91-201). These maps are  now publicly 
released  and are described in detail 
in Snowden et al. (1995). The maps cover $\sim98$ per cent of the sky
 with a resolution of 2 degrees. 
Point sources detected in the all-sky survey are included in the 
maps. The particle background, scattered solar X-ray background 
and other non-cosmic background contamination 
components have been  removed (see Snowden et al. 1994, Snowden et al. 1995).

The X-ray emission in the 1.5 keV band is  mainly
extragalactic: Hasinger et al. (1998) have resolved about 
 70 per cent of the background at these energies into 
 discrete extragalactic sources.  
 However, there is some contamination, 
 due to the poor energy resolution of the \rosat PSPC,  
 from a Galactic component. This is well-fitted with 
 a Raymond-Smith spectrum, having a temperature of $\sim$0.17 keV, 
 and may be 
associated with the Galaxy halo (see Wang \& McCray 1993, 
 Gendreau et al. 1995, Hasinger 1996, Pietz et al. 1998). 
We note  that at higher energies (3-60 keV) there is evidence for a 
even harder Galactic component with a bremsstrahlung  spectrum of 9 keV 
(Iwan et al. 1982); however, this is expected to contribute 
less than one per cent in the \rosat 1.5 keV band.  
Moreover, the 1.5 keV maps of Snowden et al. (1995)  
 show some  extended features superimposed on the 
extragalactic and the diffuse hard Galactic component (see Snowden 
et al. 1995) mainly originating from nearby supernova remnants 
(eg the North Polar Spur, the Cygnus superbubble).  
All the above local features must be subtracted before  
deriving the extragalactic X-ray dipole.

\begin{figure*}
\mbox{\epsfxsize=16cm \epsffile{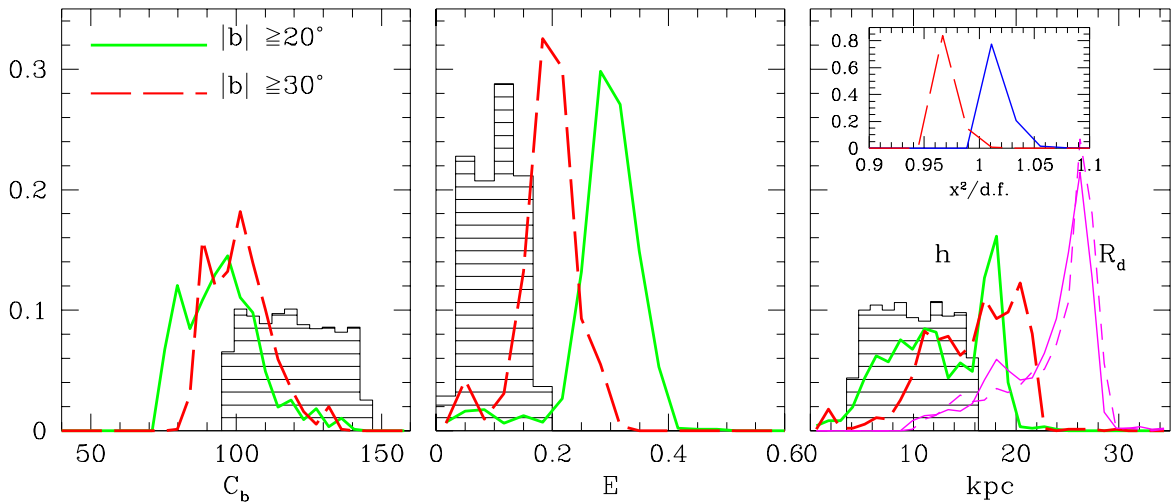}}
\caption[]{Distribution of the finite Galaxy disk model parameters. 
Continuous lines represent the output parameter distribution while the 
hatched histograms the corresponding input ones, used to start the $\chi^2$ 
minimization procedure.}
\end{figure*}

\subsection{Modelling the Galactic emission}

The derivation of a detailed Galactic emission model requires observations in 
many wavebands and is outside the scope of this paper; detailed modelling 
of the Galactic halo is discussed in Iwan et al. 
 (1982), Nousek et al. (1982) and Pietz et al. (1998). 
Here instead, we  attempt to make a rough model of the 
 Galactic contamination in our energy band. 
We model the diffuse Galactic components with 
a finite radius disk with an exponential scale height, (eg 
 Iwan et al. 1982)   
which provides a good description of the Galactic component at both 
 soft (0.75 keV) and hard energies (2-60 keV). This is given by:

\be
{\cal C}(l,b) = {\cal C}_b \left[1 + \frac{E h}{\sin|b|} 
\left(1-e^{-f(l,R_d) \tan|b|/h} \right) \right]
\ee
where $f(l,R_d)= \cos l + \sqrt{R_{d}^{2}- \sin^{2} l}$, 
with ${\cal C}$ the total X-ray intensity, ${\cal C}_b$ the average 
extragalactic component, in units of $10^{-6}$ $\rm cts ~s^{-1}~arcmin^{-2}$,
 $E$ the fraction of the total X-ray emission which 
is due to the Galaxy, $h$ and $R_d$ the disk scale height and disk radius, 
respectively both in units of 10 kpc (the galactocentric distance of the Sun).
We exclude from the fit the regions of the most apparent
extended emission features: the bulge and the North Polar Spur 
 (ie. $-40^{\circ} < b < 75^{\circ}$ and $300^{\circ} 
 \mincir l \mincir 30^{\circ}$ eg Snowden et al. 1997) as well as 
the Galactic plane strip (with $|b| < 20^{\circ}$ or $30^{\circ}$). 
 Although it is unknown whether there is some small residual  
 bulge emission outside the above excised region, 
 our rough model should provide a good first 
 order approximation to the Galactic halo emission.  
Furthermore, applying a homogeneous mean count we mask the most apparent 
"local" extragalactic features; a 
$4^{\circ}$ radius region around the Virgo cluster ($l,b \approx 287^{\circ},
75^{\circ}$) and a 10$^{\circ}$ radius region around the Magellanic clouds  
($l,b \approx 278^{\circ}, -32^{\circ}$).

In the minimization procedure we weight each pixel by $1/\sigma$, where
$\sigma=(\sigma_{I}^{2} + \sigma_{P}^{2})^{1/2}$,
with $\sigma_{I}^{2}$ the 
variance of the X-ray \rosat counts due to 
the intrinsic extragalactic fluctuations and $\sigma_{P}^{2}$ the variance 
due to Poisson count statistics. We estimate $\sigma_{I}$ 
excluding from the map the North Polar Spur and the $|b|\le 45^{\circ}$ 
regions and find $\sigma_{I} \approx 0.27 {\cal C}$. 
Starting from different initial guesses of the input 
 parameters the minimization
procedure does not reach a unique minimum, 
although the reduced $\chi^{2}$ is
around one and the output model parameters are closely clustered.
This suggests the existence of a broad and shallow minimum.
We therefore run 1000 $\chi^2$ minimizations starting from a broad range of 
initial values. These values  are centred on those that  
Iwan et al (1982) find for  the Galactic component using  
 \heao data, ie., 
$\langle h \rangle \approx 7$ kpc, $\langle R_d \rangle \approx 28$ kpc and 
$\langle E \rangle \mincir 10$ per cent; 
 while the input extragalactic contribution
is centred on $\langle {\cal C}_{b} \rangle \approx 120$, the value 
obtained from the \rosat XRB spectral fits in this band 
(eg Georgantopoulos et al. 1996). The results cluster around some preferred 
values as can be seen in figure 1, in which we show as hatched histograms the 
distribution of the input parameters  and as thick lines the best-fit output
parameter distribution. The most probable values of the fitted parameters as 
well as their standard deviation over the 1000 minimizations 
are presented in table 1. For $|b|>20^{\circ}$ we have typically that
$\chi^2\approx 33600$ for 33064 degrees of freedom and hence this 
model cannot be rejected. The Galaxy contributes a significant fraction 
($E\sim 20 - 30$ per cent) of the average total \rosat 
 1.5 keV X-ray emission.
  As discussed earlier this  Galactic component could arise 
as contamination from lower energies 
 (eg from the 0.17 keV Raymond-Smith Galactic component) 
 due to the coarse energy resolution of the \rosat PSPC. 
 Although this percentage is rather high it is not inconsistent with 
   XRB spectral fits in deep \rosat and {\it ASCA}
 pointings: 
   the excellent spectral resolution 
 {\it ASCA} spectrum of the XRB in the 0.4-10 keV 
 band (Gendreau et al. 1995) presents a steep upturn at 1 keV,
 suggesting a high contamination of the 1.5 keV \rosat band 
 from lower energy photons. 
The scale height of the finite emitting disk is $\sim 16$ kpc, 
  higher than both the value obtained 
 by Pietz et al. (1998) in the 0.75 keV \rosat band and  
 by Iwan et al (1982) in the 2-60 keV \heao band.
 Using the above best-fit model, the predicted  Galactic halo luminosity is 
 $L_x \approx 2 \times 10^{39}$ ergs $s^{-1}$. 
 The best-fit extragalactic component arises to $\sim 100\pm 11\times 10^{-6}$ 
 $\rm cts~s^{-1}~arcmin^{-2}$. This value is in excellent agreement with the 
 extrapolation of the \heao 3-60 keV data of Marshall et al. (1980)
in the \rosat band but $\sim$20 per cent lower ($\sim 1.5 \sigma$) than the 
normalization  of the extragalactic power-law component obtained from XRB 
spectral fits in deep, high galactic latitude \rosat fields 
(Hasinger 1996).
If instead we force the extragalactic contribution to the 
value obtained from the \rosat XRB fits ($\rm \sim 120 \times 10^{-6}~ 
cts~s^{-1}~arcmin^{-2}$ in this band, eg Georgantopoulos et al. 1996) 
 then the best-fit parameters are $h=4\pm 2$ kpc, $R_d=15\pm 4$ kpc and 
$E=0.28\pm 0.10$ consistent with the values of Pietz et al. (1998).
We finally note that, had we not excluded the North Polar Spur region in our 
fit, with all four parameters (${\cal C}_b, E, h, R_d$) free, we would have 
erroneously obtained a significantly higher Galactic fraction 
($E\sim 40$ per cent) as well as a larger scale height ($\sim$ 19 kpc) but with
an unacceptable fit in this case ($\chi^2_{red}\approx 1.2$). 

\begin{table}
\caption[]{Fitted Galaxy model parameters for $|b|\ge b_{lim}$.}
\tabcolsep 7pt
\begin{tabular}{ccccc} 
 $b_{lim}$ & ${\cal C}_b$ &  $E$ & $h$ (kpc) & $R_d$ (kpc) \\
$20^{\circ}$ & 95$\pm 13$ & 0.29 $\pm 0.06$ & 15 $\pm 5$ & 27$\pm 5$  \\
$30^{\circ}$ & 100$\pm 11$ & 0.20 $\pm 0.05$ & 17 $\pm 5$ & 27$\pm 5$  \\
$20^{\circ}$ & 120 (frozen) & 0.28 $\pm 0.10$ & 4 $\pm 2$ & 15$\pm 4$  \\
\end{tabular}
\end{table}

\section{XRB \rosat Dipole}

\subsection{Dipole fitting procedure}
The multipole components of the \rosat X-ray intensity are calculated
by summing moments. The dipole moment is estimated by weighing the unit 
directional vector pointing to each 40$^2$ arcmin$^2$ \rosat cell with
the X-ray intensity ${\cal C}_i$ of that cell. We normalize the dipole by the 
monopole term (the mean X-ray intensity over the sky):
\begin{equation}\label{eq:dip}
{\cal D} \equiv \frac{|{\bf D}|}{M} = \frac{\sum {\cal C}_i \hat{\bf r}_i}
{\sum {\cal C}_i}
\end{equation}
We attempt to estimate the cosmological XRB dipole by applying the above 
procedure to the \rosat counts after subtracting our best Galaxy 
model (table 1) and the regions of Galactic and "local" extragalactic X-ray 
emission. To this end we mask: (a) The Galactic plane, (b) the area dominated 
by the Galactic bulge and the North Polar Spur (see definitions in section 2.1)
and (c) a region of 10$^{\circ}$ radius around the Large Maggelanic Clouds 
(we have verified that small variations in the limits of all the above regions
do not change appreciably our results).

We use two methods to model these regions: the first consists in substituting 
the observed intensity with the mean value estimated at high galactic 
latitudes ({\em homogeneous filling procedure})  and the second based on a 
{\em spherical harmonic} extrapolation procedure (cf. Yahil et al 1986). 
Since the later is 
slightly more involved we briefly review the method which is based on
expanding the sky surface density field
$\sigma(\vartheta,\varphi)$ in real spherical harmonics:
\be
\sigma(\vartheta,\varphi) = \sum_{l,m} a_{l}^{m} Y_{l}^{m}(\vartheta,\varphi)
\ee
where $\vartheta=90^{\circ}-{\rm Gal.latitude}$, 
$\varphi={\rm Gal.longitude}$ (do not confuse the multipole $l$ with the 
Galactic Longitude). In this formulation the 
normalized dipole is defined as
\be
{\cal D}=\frac{1}{3 a_0^0} \left[\sum_{m=-1}^{1} (a^1_m)^2 \right]^{1/2}
\ee
where the factor 3 enters for consistency with the definition of 
eq.\ref{eq:dip}.
The observed, $\Sigma(\vartheta,\varphi)$, and intrinsic 
surface density field $\sigma(\vartheta,\varphi)$, are related according to:
$\Sigma(\vartheta,\varphi)~=~{\cal M}(\vartheta,\varphi)\,
\sigma(\vartheta,\varphi)$,
where the mask ${\cal M}(\vartheta,\varphi)$ takes values of 1 or 0 depending
on whether the $(\vartheta,\varphi)$ direction points in an observed or 
excluded part of the sky, respectively.
Since we are interested in recovering the dipole ($l=1$) components of
$\sigma(\vartheta,\varphi)$, the correction terms should at least involve 
the quadrupole ($l=2$) components. 
Expanding $\Sigma(\vartheta,\varphi)$ up 
to the quadrupole order and allowing for the orthogonality relation of the
Legendre polynomials, we can express the observed coefficients $A_l^m$,
in terms of the intrinsic ones, $a_l^m$, forming a $9\times 9$ matrix,
the inversion of which then gives $a_l^m$.
A more accurate procedure would entail an expansion to higher order $l$'s 
(cf. Lahav et al. 1994) but to recover a smooth underlying 
dipole structure 
(in which higher order $l$'s are negligible) we have verified, using mock 
samples, that the above procedure recovers extremely accurately the direction 
and amplitude of the true underlying dipole.

\begin{table*}
\caption[]{The \rosat Dipole Results for $|b| > b_{lim}$.}
\tabcolsep 6pt
\begin{tabular}{lcccccccc}
& mask model & $b_{lim}$ & $l^{\circ}$ & $b^{\circ}$ & 
$\delta\theta_{\rm CMB_{\odot}}^{\circ}$ &
$\delta\theta_{\rm CMB_{LG}}^{\circ}$ & ${\cal D}$ & sky masked \\ \\
{\em raw counts} & {\em Homogeneous} &20 &342.7 &  20.4 & 67.6 & 59.5 & 0.050 
& 36\% \\
{\em $\;\;\;\;\;\;\;\;$ ''} & {\em Sph.Harm.} & 20& 343.0 &  11.0 & 74.5 & 
63.8 & 0.092 & 36\% \\

{\em Galaxy, North Polar Spur} & {\em Homogeneous} &  20& 280.3  & 
29.1 & 21.2    &  3 &0.011 & 47\%\\
{\em \& Magellanic Clouds excluded} & {\em Sph.Harm.} &  20 & 318.3  & 
20.1 & 51.1  &  38.5 & 0.037 & 47\% \\ 

{\em $\;\;\;\;\;\;\;$ ''} & {\em Homogeneous} &  30& 290.9  & 
41.6 & 19.8  & 16.1 & 0.010 & 63\%\\
{\em $\;\;\;\;\;\;\;$ ''} & {\em Sph.Harm.} &  30 & 313.5  & 
14.9 & 51.7  & 36.7 & 0.049 & 63\% \\
\end{tabular}
\end{table*}

\subsection{Dipole Results}
In table 2 we present our main results for different treatment of the data.
It is evident that when using the raw {\small ROSAT} data, the dipole points
roughly towards the Galactic centre (in agreement with 
Kneissl et al 1997). However, when we exclude both the Galaxy and the North 
Polar Spur, the measured dipole is in much better directional agreement with
the CMB dipole. For the {\em homogeneous filling} method we find
$\delta\theta_{cmb} < 20^{\circ}$ excluding the Galactic plane
below $|b|=20^{\circ}$ or $|b|=30^{\circ}$. For the {\em 
spherical harmonic} method the misalignment angle is larger, 
$\delta\theta_{cmb} > 39^{\circ}$.
The Virgo cluster, being quite near and very bright in X-rays, could 
contribute significantly to the measured dipole. 
Excluding an area of 4$^{\circ}$ radius around the Virgo cluster
($l,b \approx 287^{\circ},75^{\circ}$), which is very apparent in the X-ray 
map, we find that it is responsible for $\sim 20$ per cent 
 of the total dipole, while the dipole misalignment angle with the CMB 
increases by $\sim 15^{\circ}$.

However, it should be expected that many Galactic sources, probably 
dominating the higher \rosat counts, are still present in 
the data and could affect the behaviour of the extragalactic XRB dipole. We
therefore present in figure 2 the misalignment angle between
the \rosat and CMB dipoles as well as the normalized dipole amplitude,
${\cal D}$, as a function of the \rosat upper count limit 
(${\cal C}_{up}$).
The errorbars have been estimated by using the different Galactic model 
parameters resulting from the $\chi^2$ fits of section 2.1 and presented in
figure 1.
We do find that our main results are very robust in such variations of the 
Galactic model.
The two methods, used to mask the excluded regions, give consistent 
dipole results for ${\cal C}_{up} \mincir 140$ (which cover $\sim$ 97 
per cent of the unmasked sky).
For this limit the \rosat-CMB dipole misalignment angle is
$\mincir 26^{\circ}$ and $33^{\circ}$ for the {\em homogeneous filling} 
and {\em spherical harmonics} methods respectively. It is evident that the 
\rosat-CMB dipole misalignment angle increases substantially when we 
include the few higher intensity cells, before however the Virgo cluster 
(which enters at ${\cal C}_{up} \magcir 200$ counts) starts reducing again 
the misalignment angle, as can be clearly seen in figure 2. 
The interpretation that the high intensity (${\cal C}\magcir 140$) cells are 
associated with Galactic sources is supported by the fact that when we 
include these few cells the resulting dipole direction moves 
towards the Galactic centre. We therefore consider as our best estimate
of the XRB dipole its value at ${\cal C}_{\rm up} \approx 140$, for which 
{\em both
methods used to model the masked areas agree and the XRB-CMB dipole 
misalignment angle is minimum}.

To take into account all possible sources of uncertainty, we
use a Monte-Carlo simulation approach in which we vary all the model 
parameters within their range of validity. Using 6000 dipole realizations 
to take into account (a) the 
uncertainties of the Galactic model subtracted from the raw counts, (b)  
the different methods used to mask the excluded sky regions (c) the 
different galactic latitude limits and (d) variations of the excision radii
around the bulge and the North Polar Spur we conclude that the 
XBG \rosat dipole has:
$$ {\cal D}_{\rm XRB} \approx 0.017 \pm 0.008 \;\;\;\;\; (l,b) 
\approx (286^{\circ}, 22^{\circ}) \pm 19^{\circ} $$
which deviates from the CMB dipole directions in the heliocentric
and Local Group frames by $\delta\theta_{\rm CMB_{\odot}} \sim 
30^{\circ}$ and $\delta\theta_{\rm CMB_{LG}}\sim 10^{\circ}$, respectively. 
It is interesting that 
the \rosat dipole is nearer to the Local Group frame CMB dipole direction.
Our  results are consistent with the \heao (2-10 keV) dipole 
(Shafer \& Fabian
1983) which points in a  similar direction ($282^{\circ}, 30^{\circ}$), albeit 
with a larger uncertainty,  
but has a lower amplitude: ${\cal D}_{\rm HEAO-1} \sim 0.005$.
 
\begin{figure}
\mbox{\epsfxsize=11cm \epsffile{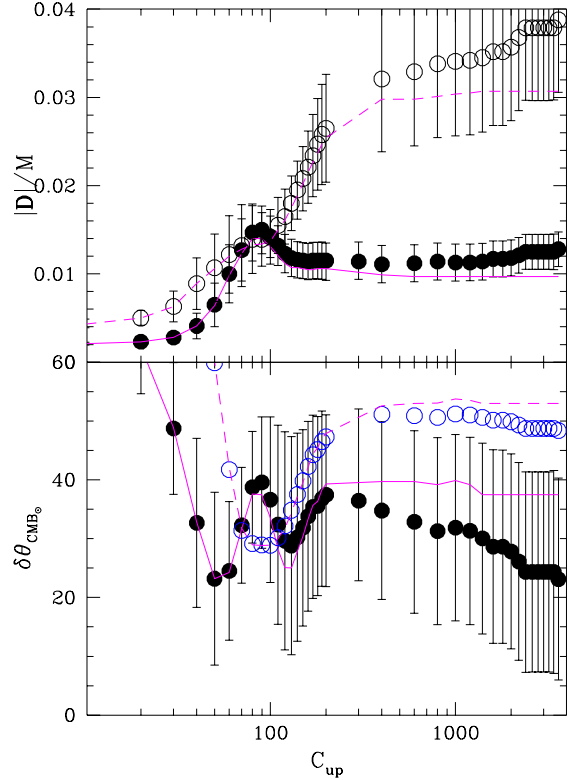}}
\caption[]{Dipole results as a function of upper count limit, ${\cal C}_{up}$.
The filled and open points correspond to the {\em homogeneous fill} and {\em
Spherical harmonics} mask methods respectively. The scatter correspond to 
the variations of the Galaxy model, as given by all the different $\chi^2$
fits (see section 2.1).
The corresponding solutions, excluding the Virgo cluster, are shown as the
continuous and dashed lines.}
\end{figure}

\subsection{Interpretation}
The motion of the Sun with respect to an isotropic radiation background
produces a dipole in the radiation intensity according to: 
\be 
\frac{\delta {\cal C}}{\langle {\cal C} \rangle}=(3 + \alpha) 
V_{\odot} \cos \theta /c
\ee
where $\alpha$ is the spectral index of the radiation (${\cal C} \propto 
\nu^{-\alpha}$. For the 1.5 keV \rosat band we have $\alpha \sim 0.4$
 (Gendreau et al. 1995).
If the \rosat dipole was totally due to the motion of the Sun with
respect to the XRB (Compton-Getting effect) then we would obtain a solar 
velocity with respect to the XRB of $V_{\odot}=1300 \pm 600$ km/sec, which
should be compared with $V_{\odot}=369$ km/sec with respect to the 
CMB (Lineweaver et al 1996).

We therefore verify that the observed XRB \rosat dipole 
cannot be only due to the Compton-Getting effect but it is significantly
{\em contaminated} by the dipole produced by X-ray emitting sources that 
trace the large-scale structure. 
We estimate the large-scale dipole component 
of the XRB by subtracting from the \rosat map the expected
Compton-Getting dipole and we obtain a dipole with ${\cal D}_{\rm LSS} 
\approx 0.9 {\cal D}_{\rm XRB}$ pointing towards $(l,b) 
\sim (284^{\circ},18^{\circ})$.
This is in general agreement with Lahav et al (1997) who
found that the two effects, contributing to the XRB dipole, are of the same 
order of magnitude. 

\section{Conclusions}
We have estimated the dipole of the diffuse 1.5 keV X-ray background using 
the \rosat all-sky survey maps of Snowden et al (1995). We have first   
subtracted from the \rosat counts the diffuse Galactic emission  
 (the halo and bulge components) 
as well as local extended features such as the North Polar Spur.
The Galactic halo model used is that of a finite radius disk model with 
an exponential scale height (eg Iwan et al 1982). 
 The mean Galactic X-ray component is $\sim
20 -30$ per cent of the background and the scale height and radius are  
$\sim 16 \pm 5$ and 27$\pm 5$ kpc respectively. We model the excluded regions 
by either homogeneously ``painting'' these 
regions with the mean X-ray count (derived after subtracting the Galaxy) or 
using a spherical harmonic expansion of the X-ray surface intensity field. 

We estimate that  the \rosat XRB dipole is pointing towards 
 $(l,b) \approx (286^\circ, 22^\circ) \pm 19$, within $\sim 30^{\circ}$
of the CMB direction and having a normalized amplitude of $\sim 1.7$ per cent.
The dipole direction is in agreement with previous estimates in hard X-rays 
(Shafer \& Fabian 1982) but the positional errors have now been improved.   
We also find that the two effects expected to contribute to the XRB,
ie., the Compton-Getting effect and the anisotropy due to X-ray sources
tracing the large-scale structure, are of the same order of magnitude, in 
general agreement with the predictions of Lahav et al (1997).
However, the latter dominates having ${\cal D}_{\rm LSS} \sim 0.9 
{\cal D}_{\rm XRB}$.
Finally we estimate that the nearest cluster, Virgo, contributes $\sim 20$
 per cent to the total measured XRB dipole.

\section*{Acknowledgments}
We thank the referee, Dr. Marie Treyer, for her helpful comments and 
suggestions.


{\small 
\begin{thebibliography}{}
\bibitem[]{}Boldt E., 1987, Phys. Rep., 146, 215
\bibitem[]{}Compton A., Getting I., 1935, Phys. Rev., 47, 817
\bibitem[]{}Gendreau. K. et al. 1995, PASJ, 47, L5
\bibitem[]{}Georgantopoulos, I., Stewart, G.C., Shanks, T., 
 Boyle, B.J., Griffiths, R.E., 1996, MNRAS, 280, 276  
\bibitem[]{}Hasinger, G., in IAU Symposium 168, eds Kafatos M. \& Kondo Y., 
p. 245, Kluwer
\bibitem[]{}Hasinger, G., Burg, R., Giaconni, R., Schmidt, M., 
Trumper, J., Zamorani, G., 1998, AA, 329, 482
\bibitem[]{}Iwan D., Marshall F.F., Boldt E.A., Mushotzky R.F., Shafer R.A., 
\& Stottlemyer A., 1982, ApJ, 260, 111
\bibitem[]{} Jahoda K., 1993, Adv.Space Res., 13, (12)231
\bibitem[]{}Kneissl R., Egger R., Hasinger G., Soltan A.M. \& Trumper J.,
1997, AA, 320, 685
\bibitem[]{}Lahav O., Edge, A.C., Fabian, A.C., Putney A., 1989, MNRAS, 
238, 881
\bibitem[]{}Lahav O., Fisher, K.B., Hoffman, Y., Schraf, C.A. \& Zaroubi, S.,
1994, ApJ, 423, L93
\bibitem[]{}Lahav O., Piran T. \& Treyer, M., 1997, MNRAS, 284, 499
\bibitem[]{}Lineweaver C.H. et al, 1996, ApJ, 470, 38
\bibitem[]{}Marshall, F.E., Boldt, E.A., Holt, S.S., Miller, R.B., 
Mushotzky, R.F., Rose, L.A., Rothschild, R.E., Serlemitsos, P.J., 1980,
ApJ, 235, 4 
\bibitem[]{}Miyaji T., Boldt E., 1990, ApJ, 353, L3
\bibitem[]{}Nousek, J.A., Fried, P.M., Sanders, W.T., Kraushaar, W.L., 1982, 
 ApJ, 258, 83
\bibitem[]{}Plionis M. \& Kolokotronis V., 1998, ApJ, 500, 1 
\bibitem[]{} Pietz, J., Kerp, J., Kalberla, P.M.W., Burton, W.B., Hartmann, 
D., Mebold, U., 1998, AA, 332, 55  
\bibitem[]{}Protheroe, R.J., Wolfendale, A.W., Wdoczyk, J., 1980, MNRAS,
 192, 445
\bibitem[]{}Shafer R.A., Fabian A.C., in Abell G.O., Chincarini G., eds, IAU
Symp. No 104, Early Evolution of the Universe and its Present Structure. 
Kluwer, Dordrecht, p.33
\bibitem[]{}Snowden S.L., McCammon, D., Burrows, D. N., Mendelhall, J. A., 
1994, ApJ, 424, 714
\bibitem[]{}Snowden S.L. et al, 1995, ApJ, 454, 643
\bibitem[]{}Wang, Q.D., McCray, R., ApJ, 409, L37
\bibitem[]{}Yahil A., Walker, D., Rowan-Robinson, M., 1986, ApJ, 301, L1
\end{thebibliography}
}
\end{document}